\documentclass[aps,prl,twocolumn,showpacs,amssymb,nofootinbib]{revtex4}
\usepackage{graphicx}

\begin{document}
\title{Exact Quantization of Einstein-Rosen Waves Coupled to Massless Scalar Matter}

\author{J. Fernando \surname{Barbero G.}}
\email[]{fbarbero@iem.cfmac.csic.es} \affiliation{Instituto de
Estructura de la Materia, C.S.I.C., Serrano 123, 28006 Madrid,
Spain}
\author{I\~naki  \surname{Garay}}
\email[]{igael@iem.cfmac.csic.es} \affiliation{Instituto de
Estructura de la Materia, C.S.I.C., Serrano 123, 28006 Madrid,
Spain}
\author{Eduardo J. \surname{S. Villase\~nor}}
\email[]{ejsanche@math.uc3m.es} \affiliation{Departamento de
Matem\'aticas, Escuela Polit\'ecnica Superior, Universidad Carlos
III de Madrid, Avda. de la Universidad 30, 28911 Legan\'es, Spain}
\affiliation{Instituto de Estructura de la Materia, C.S.I.C.,
Serrano 123, 28006 Madrid, Spain}

\date{May 6, 2005}

\begin{abstract}
We show in this letter that gravity coupled to a massless scalar
field with full cylindrical symmetry can be exactly quantized by
an extension of the techniques used in the quantization of
Einstein-Rosen waves. This system provides a useful testbed to
discuss a number of issues in quantum general relativity such as
the emergence of the classical metric, microcausality, and large
quantum gravity effects. It may also provide an appropriate
framework to study gravitational critical phenomena from a quantum
point of view, issues related to black hole evaporation, and the
consistent definition of test fields and particles in quantum
gravity.
\end{abstract}


\maketitle

Symmetry reductions of general relativity have been used as model
systems to extract information about quantum gravity. They usually
allow the discussion of specific problems without the difficulties
present in the full theory. Some of the most popular choices in
this regard (Bianchi models) have only a finite number of degrees
of freedom and, hence, are not suitable to address some of the
more nagging questions posed by the study of quantum gravity
(diffeomorphism invariance or issues related to the presence of an
infinite number of local degrees of freedom such as perturbative
non-renormalizability). Fortunately there are other symmetry
reductions that retain these features while still being exactly
solvable both at the classical and quantum levels. Chief among
them are the Einstein-Rosen waves \cite{ER,Ku}, obtained by
requiring that space-time metrics have two commuting, spacelike,
and hypersurface orthogonal Killing vector fields (one
translational and the other rotational). This model has been
extensively studied in the past \cite{AP,AM} and some intriguing
results have been derived, in particular the appearance of
unexpected large quantum gravity effects \cite{Ash} and a detailed
picture of the emergence of the causal structure of space-time in
the classical limit \cite{Nos1,Nos2}. An improvement that would
increase the usefulness of this system as a toy model for quantum
gravity would be the coupling of matter. The availability of a
\textit{solvable} model with matter would open up a host of
interesting possibilities deserving a careful investigation. We
show in this letter that such a model exists and discuss how it
can be \textit{exactly} quantized. Specifically we will consider
here the quantization of Einstein-Rosen waves coupled to a
cylindrically symmetric massless scalar field.

To our knowledge this system was first discussed from a classical
point of view by Chandrasekhar \cite{Chandra} who showed that a
full solution to the Einstein field equations can be found in this
case. The specific form of this solution suggests that a
Hamiltonian treatment of the system would lead to a description
very similar to the one found by Ashtekar, Pierri, and Varadarajan
\cite{AP,AM}  for pure gravity in the asymptotically flat case.
This is a strong indication that the model is amenable to
quantization by a suitable extension of known techniques. The main
point of this letter is to show that this is indeed the case.

The possible applications of such a model are manifold and can be
classified in several different categories. First of all there is
the issue of extracting information about geometry in quantized
gravity. We do not expect to find here the kind of precise
geometric information offered by loop quantum gravity in the form
of geometric observables such as areas or volumes. Nevertheless
our approach gives some indications about the validity of a metric
description in the realm of quantum gravity. This has already been
considered for pure gravity by studying expectation values of
metric components. Here we propose to follow a different
philosophy; instead of obtaining some approximate semiclassical
metric by taking expectation values of a metric operator in some
quantum state we can use the scalar field and its particle-like
excitations to explore space-time geometry operationally, much in
the same way as one uses the geodesics followed by test particles
to understand the geometrical features of a given space-time
metric. The availability of an external probe allows us to discuss
the microcausality of the model both from the perspective of the
gravitational and scalar degrees of freedom. The agreement between
both points of view, that we discuss later, is a clear indication
of the usefulness of the present approach in the discussion of
quantum gravitational effects and supports the results already
obtained for the purely gravitational case.

A second set of questions that can possibly be addressed within
this framework \textit{in the quantum regime} are related to
critical phenomena in gravitational collapse and problems in black
hole physics\footnote{This would require dropping the radial
asymptotic flatness condition. By doing this we can have the
self-similar solutions needed to discuss critical collapse
\cite{Wang} or escape the conclusions of \cite{BCM} about the
absence of compact trapped surfaces in the asymptotically flat
case.}. These issues have been recently considered by Wang in his
study of critical collapse of a cylindrically symmetric scalar
field in four dimensions \cite{Wang}. This system displays some
rich and non-trivial behavior, in particular, the possibility of
forming solutions with future, spacelike singularities by the
collapse of massless scalar matter and the appearance of a
critical metric separating solutions with different singular
behavior. Notice that having an exact solution, backreaction
effects are automatically taken into account without any
approximation. Finally we want to point out other possible uses of
this model such as the discussion of the validity of the usual
perturbative schemes in quantum gravity, the development of new
ones, the discussion of issues in QFT in curved spacetimes, and
the application to other useful symmetry reductions of these type
(Gowdy models) that are similar to Einstein-Rosen waves and,
hence, can also be solved after coupling massless scalar fields.

Our starting point is the four dimensional action for a massless
scalar $\Phi_s$ coupled to gravity with cylindrical symmetry
\begin{eqnarray*}
& &^4S=\frac{1}{16\pi G_N}\int_{\mathcal{M}\times
I}\!\!\!\!d^4x\,\sqrt{|^4\!g|}\Big[R-\frac{1}{2}
{^4\!g^{ab}}\nabla_a\Phi_s\nabla_b\Phi_s\Big]\quad\nonumber\\
& &\hspace{7mm}+\frac{1}{8\pi G_N}\int_{\partial(\mathcal{M}\times
I)}\!\!\!\!d^3x\,(\sqrt{|^3\!h|}K-\sqrt{|^3\!h^0|}K^0)\,.
\end{eqnarray*}
Here we have included the surface terms necessary to have a
well-defined variational principle, $I\equiv[z_1,z_2]$ is a closed
interval in the direction of the translational Killing vector
$\partial_z$, and $K$, $K^0$ are the extrinsic curvatures of the
boundary defined by the dynamical metric $^4g_{ab}$ and a fiducial
metric $^4g^0_{ab}$ that we choose as Minkowski in the following
(we denote the induced metrics on the boundary as $^3h_{ab}$ and
$^3h_{ab}^0$). The Geroch formalism (and a subsequent conformal
transformation) allows us to reduce the previous action to the
following three dimensional one by taking advantage of the
translational symmetry
\begin{eqnarray}
^3S\!\!&=&\!\!\frac{1}{16\pi
G_3}\int_{\mathcal{M}}\!\!\!\!d^3x\,\sqrt{|g|}\Big[\,^3\!
R-\frac{1}{2}g^{ab}\nabla_a\phi_g\nabla_b\phi_g\label{001}\\
&&\hspace{-.9cm}-\frac{1}{2}g^{ab}\nabla_a\phi_s\nabla_b\phi_s\Big]+\frac{1}{8\pi
G_3}\int_{\partial\mathcal{M}}\!\!\!\!\!\!d^2x\,\big[\sqrt{|h|}K-\sqrt{|h^0|}K^0\big]\,.\nonumber
\end{eqnarray}
Here $g_{ab}$ is a 3-dimensional metric and $^3\!R$ the
corresponding scalar curvature, $\phi_g$ is the scalar field that
encodes the local gravitational degrees of freedom of the model
\cite{AP}, $\phi_s$ is the massless matter scalar field in three
dimensions, and $G_3$ is the gravitational constant per unit
length along the symmetry axis (with dimensions of inverse energy;
in the following we choose units such that $\hbar=c=8G_3=1$). The
integration is extended to a 3-manifold $\mathcal{M}$ with
boundary $\partial\mathcal{M}$ with the appropriate topology. At
this point it could be argued that the inclusion of the massless
scalar is a rather trivial addition to the system because it just
plays the role of an extra field of the same type of the
gravitational scalar already present in the 2+1 dimensional
description of Einstein-Rosen waves. However this is the most
important and unexpected feature of (\ref{001}) because
\textit{both} the gravitational and matter degrees of freedom are
described by the same type of term in the three-dimensional
Lagrangian in spite of their very different meaning in the
original action, the Geroch reduction, and the conformal
transformation used to arrive at (\ref{001}). It is also striking
that they couple only through the metric and not directly (there
are no cross terms).

As we are interested in the quantization of the system it is
necessary to obtain the Hamiltonian corresponding to (\ref{001}).
Although the final answer turns out to be quite simple it is not
completely obvious, and it has some surprising features, so we
provide some details on its derivation. To this end we choose a
foliation of $\mathcal{M}$ with timelike unit normal $n^a$, a
radial unit vector $\hat{r}^a$, and denote as $\sigma^a$ the
azimutal, hypersurface orthogonal, Killing field (notice that this
is not a unit vector). We further introduce two additional vector
fields $t^a$ and $r^a$ defined as $t^a=N n^a+N^r \hat{r}^a$ and
$r^a=e^{\gamma/2}\hat{r}^a$, where $N$ is the lapse function,
$N^r$ the radial shift, and $\gamma$ is an additional field. It is
possible to find conditions that ensure that $t^a$, $r^a$, and
$\sigma^a$ are coordinate vectors. If we define
$\partial_{\sigma}\equiv\sigma^a\nabla_a$, $\partial_{r}\equiv
r^a\nabla_a$, and $\partial_{t}\equiv t^a\nabla_a$ these are the
following:
\begin{eqnarray*}
&&\partial_{\sigma}N=\partial_{\sigma}N^r=\partial_{\sigma}\gamma=0\,;\,[\sigma,\hat{r}]^a=[\sigma,n]^a=0\,;\\
&& n^a\partial_r N+\hat{r}^a(\partial_rN^r-\partial_t
e^{\gamma/2})+Ne^{\gamma/2}[\hat{r},n]^a=0.
\end{eqnarray*}
Writing the metric as
$g_{ab}=-n_an_b+\hat{r}_a\hat{r}_b+\frac{1}{R^2}\sigma_a\sigma_b$
(with $R^2\equiv g_{ab}\sigma^a\sigma^b$) we obtain the line
element in these coordinates
\begin{eqnarray*}
ds^2=(N^{r2}-N^2)dt^2+2e^{\gamma/2}N^rdtdr+e^{\gamma}dr^2+R^2d\sigma^2.
\end{eqnarray*}
The action can be rewritten now as
\begin{eqnarray*}
^3S&=&\int_{t_1}^{t_2}\!\!dt\int_0^{\tilde{r}}\!\!dr
\Big\{Ne^{-\gamma/2}(\gamma^{\prime}R^{\prime}-2R^{\prime\prime})\hspace{1cm}\label{006}\\
&&\hspace{-.3cm}-\frac{1}{N}(e^{\gamma/2}\dot{\gamma}-2N^{r\prime})
(\dot{R}-e^{-\gamma/2}N^rR^{\prime})+\nonumber\\
&&
\hspace{-.3cm}+\frac{R}{2N}\Big[e^{\gamma/2}\dot{\phi}^2_g-2N^r\dot{\phi}_g{\phi}_g^{\prime}+
e^{-\gamma/2}(N^{r2}-N^2){\phi}_g^{\prime2}\Big]\nonumber\\
&&
\hspace{-.3cm}+\frac{R}{2N}\Big[e^{\gamma/2}\dot{\phi}^2_s-2N^r\dot{\phi}_s{\phi}_s^{\prime}+
e^{-\gamma/2}(N^{r2}-N^2){\phi}_s^{\prime2}\Big]\Big\}\nonumber\\
&&\hspace{-.3cm}+2\int_{t_1}^{t_2}\!\!dt(Ne^{-\gamma/2}R^{\prime}-1)\nonumber\,,
\end{eqnarray*}
where we have denoted $\partial_t$ with a dot, $\partial_r$ with a
prime. The Hamiltonian when we take $\tilde{r}\rightarrow\infty$
is
\begin{eqnarray*}
H\!\!&=&\!\!\int_0^{\infty}\!\!\!\!dr
\Big\{N^re^{-\gamma/2}\Big[p_RR^{\prime}-p^{\prime}_{\gamma}+p_{\gamma}\gamma^{\prime}+
\phi^{\prime}_gp_g+\phi^{\prime}_sp_s\Big]\nonumber\\
&&\hspace{8mm}+Ne^{-\gamma/2}
\Big[2R^{\prime\prime}-\gamma^{\prime}R^{\prime}-p_Rp_{\gamma}+\frac{1}{2R}p_g^2\label{008}\\
&&\hspace{8mm}+\frac{R}{2}\phi_g^{\prime2}+\frac{1}{2R}p_s^2+\frac{R}{2}\phi_s^{\prime2}\Big]\Big\}
+2(1-e^{-\gamma_{\infty}/2})\,,
\end{eqnarray*}
where $p_R$, $p_{\gamma}$, $p_g$, and $p_s$ are the momenta
canonically conjugate to $R$, $\gamma$, $\phi_g$, and $\phi_s$
respectively, $\gamma_{\infty}\equiv
\lim_{r\rightarrow\infty}\gamma(r)$, and the fall-off of the
fields, that ensures asymptotic flatness in 2+1 dimensions and
implies $N\rightarrow1$ and $R^{\prime}\rightarrow1$,  is the one
used in \cite{AP}. All fields are chosen to be regular in the
axis. From the previous expression the Hamiltonian of the system
and the constraints can be immediately read. To proceed ahead we
fix the gauge with the conditions $R(r)=r$ and $p_{\gamma}(r)=0$
(the same as in the absence of matter). It is straightforward to
show that they are admissible. After fixing the gauge and solving
the constraints we get
\begin{eqnarray*}
\gamma(R)=\frac{1}{2}\int_0^{R}\!\!\!dr\,
r\Big[\phi_g^{\prime2}+\frac{p_g^2}{r^2}+\phi_s^{\prime2}+\frac{p_s^2}{r^2}\Big],
\end{eqnarray*}
the three dimensional line element can be written as
\begin{equation}
ds^2=e^{\gamma}[-e^{-\gamma_{\infty}}dt^2+dR^2]+R^2d\sigma^2\label{009}
\end{equation}
and the reduced Hamiltonian is
$$H=2(1-e^{-\gamma_{\infty}/2}).$$
This is a function of the sum of the Hamiltonians for two massless
cylindrically symmetric fields evolving in a fictitious
Minkowskian background. For every solution to the field equations
$\gamma_{\infty}$ is a constant of motion. Taking advantage of
this we can introduce an auxiliary, \textit{solution-dependent},
time variable as in \cite{AP} defined according to
$T=e^{-\gamma_{\infty}/2}t$, that allows us to simplify the form
of the field equations to get
\begin{eqnarray}
&&\hspace{-1cm}\partial^2_{T}\phi_g-\phi_g^{\prime\prime}-\frac{1}{R}\phi_g^{\prime}=0,
\quad\partial^2_{T}\phi_s-\phi_s^{\prime\prime}-\frac{1}{R}\phi_s^{\prime}=0\,.\label{011}
\end{eqnarray}
Equations (\ref{011}) describe two massless, cylindrically
symmetric scalar fields in 2+1 dimensions. Classically this is a
time redefinition that amounts to a change of the coordinate $t$;
once we pick a certain solution to (\ref{011}) we can choose to
write (\ref{009}) either in terms of $t$ or $T$. Quantum
mechanically the situation is more complicated because the
evolution of wave packets generically involves the superposition
of Hilbert space vectors with energy dependent phases so a change
in the functional form of the energy completely changes the
evolution of the states. It is very important to notice that the
form of the Hamiltonian means that the model \textit{is not free}.
The two fields that appear are coupled in a non trivial way.

In order to quantize the system we define field and momenta
operators $\hat{\phi}_{g,s}(R)$, $\hat{p}_{g,s}(R)$ satisfying the
commutation relations
$[\hat{\phi}_{g,s}(R),\hat{p}_{g,s}(R^{\prime})]=i\delta(R,R^{\prime})$
and introduce creation and annihilation operators as usual
according to
\begin{eqnarray*}
&&
\hspace{-11mm}\phi_{g,s}(R)=\!\!\frac{1}{\sqrt{2}}\!\!\int_0^{\infty}\!\!\!\!\!dk
\,J_0(Rk)[a_{g,s}(k)+a_{g,s}^{\dagger}(k)]\,,\\
&&
\hspace{-11mm}p_{g,s}(R)=\!\!\frac{iR}{\sqrt{2}}\!\!\int_0^{\infty}\!\!\!\!\!
dk\,kJ_0(Rk)[a_{g,s}^{\dagger}(k)-a_{g,s}(k)]\,,
\end{eqnarray*}
with non-zero commutators given by
\begin{eqnarray*}
&&[a_g(k),a^{\dagger}_g(q)]=\delta(k,q),\,[a_s(k),a^{\dagger}_s(q)]=\delta(k,q).\quad\quad\label{015}
\end{eqnarray*}
These operators are defined in a Hilbert space built as a tensor
product of two Fock spaces $\mathcal{H}_g$ and $\mathcal{H}_s$,
$\mathcal{H}=\mathcal{H}_g\otimes\mathcal{H}_s$ with a vacuum
state $|\Omega\rangle=|0\rangle^g\otimes|0\rangle^s$ defined in
terms of the vacua annihilated by $a_{g,s}(k)$. States with a
fixed number of quanta of ``gravitational" or ``scalar" type are
obtained by repeated action of the corresponding creation
operators $|k\rangle_{g,s}\equiv
A_{g,s}^{\dagger}(k)|\Omega\rangle$, where we have written
$A^{\dagger}_{g}(k)\equiv a^{\dagger}_{g}(k)\otimes\mathbb{I}_s$,
$A^{\dagger}_{s}(k)\equiv \mathbb{I}_g\otimes a^{\dagger}_{s}(k)$.

The quantum Hamiltonian in $\mathcal{H}$ is
\begin{eqnarray*}
\displaystyle
\hat{H}\!=\!2\,\Big(1-\exp\!\Big[\!-\!\frac{1}{2}\!\displaystyle\int_0^{\infty}\!\!\!\!\!\!\!dk\,k[A_g^{\dagger}(k)A_g(k)
+A_s^{\dagger}(k)A_s(k)]\Big]\!\Big) .\nonumber
\end{eqnarray*}
We have normal ordered the exponent to remove the zero point
energy of the vacuum. This Hamiltonian is a non-linear and bounded
function of the sum of the Hamiltonians for two massless,
cylindrically symmetric scalar fields in 2+1 dimensions, $H_0^g$
and $H_0^s$. It is an observable of the system (the energy) and
the generator of time evolution in the time variable $t$ (from
$t_1$ to $t_2$)
\begin{equation}
U(t_2-t_1)=\exp\Big[-2i(t_2-t_1)\big(1-e^{-\frac{1}{2}[H_0^g+H_0^s]}\big)\Big].\label{016}
\end{equation}
It is important to realize at this point that this is the physical
evolution. The free Hamiltonians $H_0^g$ and $H_0^s$ are indeed
observables but are not directly related to the time evolution of
the system. It is necessary to take this fact into account in the
search for semiclassical states because coherent states should
display a classical behavior under the evolution given by
(\ref{016}) rather than under the one that would be defined by the
``free" Hamiltonian $H_0^g+H_0^s$. The expectation value of the
field and momenta operators should evolve in $t$ according to the
classical field equations in terms of $t$; notice that they are
not (\ref{011}).

The unitary evolution given by $U(t)$ defines the $S$-matrix of
the system (the $S$-matrix in QFT is basically the evolution
operator in the limit $t\rightarrow\infty$). Its matrix elements
in n-particle states can be computed in a straightforward way
because they are eigenstates of $H_0^g$ and $H_0^s$. The only
non-zero matrix elements on states with a definite number of both
quanta (i.e. gravitational and matter) are those connecting state
vectors with the same number of particles of each type; hence
there is no conversion of quanta of one type into the other. It is
necessary at this point to stress that the split of the Hilbert
space as a tensor product of two Hilbert spaces should not
immediately lead us to interpret one of them as ``gravitational"
and the other as ``matter"; in fact the classical metric depends
on both the gravitational scalar and the matter scalar and
semiclassical approximations of it obtained by computing
expectation values of a metric operator would also depend on both
the ``gravitational" and the ``matter" part of the state. In this
sense a vector such as $|0\rangle^g\!\otimes|\Phi\rangle^s$ should
not be interpreted as an approximate description of a pure matter
state $|\Psi\rangle^s$ on some quantum approximation of the
Minkowski metric. In fact the quantum state that most closely
resembles the Minkowski metric is the vacuum $|\Omega\rangle$. By
the way, this is the only coherent state of the system that we
know under the evolution given by (\ref{016}).

The fact that the $S$-matrix on n-particle states can be found in
a straightforward way on the generalized orthonormal basis
$|p\rangle^g\!\otimes|q\rangle^s$ and its diagonal character does
not mean that interesting quantum information on the system cannot
be obtained. Quite on the contrary some significant features of
quantum spacetime geometry can be seen. We will concentrate here
on the discussion of microcausality. Though we have looked at this
problem somewhere else \cite{Nos1,Nos2} the inclusion of scalar
matter in the model allows us to consider the issue both from the
point of view of the metric scalar and the matter scalar. As in
previous work we concentrate on the vacuum expectation value of
the commutator of the fields evolved with (\ref{016}) in the
Heisenberg picture
\begin{eqnarray}
&&\hspace{-4mm}\langle\Omega|[\hat{\phi}_g(R^{\prime},t^{\prime}),
\hat{\phi}_g(R,t)]|\Omega\rangle\!\!=\!\!
\langle\Omega|[\hat{\phi}_s(R^{\prime},t^{\prime}),\hat{\phi}_s(R,t)]|\Omega\rangle\nonumber\\
&&=-\frac{i}{2}\int_0^{\infty} \!\!\!dk\,
J_0(R^{\prime}k)J_0(Rk)\sin{[(t^{\prime}-t)E(k)]}\,, \label{017}
\end{eqnarray}
with $E(k)=2(1-e^{-k/2})$.  As can be seen the commutator on the
vacuum for both types of fields is \textit{exactly} the same. The
microcausality of the model as described by the gravitational and
matter scalars coincide. As we showed in \cite{Nos2}, we can see
from (\ref{017}) the emergence of the sharp and well defined
cylindrical light cone structure corresponding to the quantization
of a cylindrical massless scalar field in a 2+1-dimensional
Minkowskian background. This happens in the limit when spatial
distances and time intervals are large in comparison with the
length scale $\hbar G_3$. Finally it is interesting to point out
that even though the diagonal matrix elements of the commutator of
$\hat{\phi}_g(R_2,t_2)$ and $\hat{\phi}_s(R_1,t_1)$ are zero, the
non-diagonal ones are generically different from zero for $t_2\neq
t_1$. This is a clear indication that both fields interact in a
non-trivial way.

The availability of a matter field allows us to explore how the
quantization of gravity reflects on the geometric properties of
spacetime by using its particle-like excitations as quantum test
particles. Though the details of this will appear elsewhere we
want to explain here how it can be done. A possible approach to
the problem is to interpret
$\langle\Omega|\hat{\phi}_s(R_2,t_2)\hat{\phi}_s(R_1,t_1)|\Omega\rangle$
as the probability amplitude for a particle created at the radial
distance $R_1$ in the instant $t_1$ to be found at $(R_2,t_2)$. As
in ordinary QFT in Minkowski spacetime this interpretation is only
approximate (i.e. valid only above a certain distance scale)
because the $R$-dependent states $\hat{\phi}_s(R)|\Omega\rangle$
do not constitute an orthonormal basis. One can, however,
introduce the analogous of the Newton-Wigner localized states for
this system, define one-particle wave functions depending on the
radial coordinate, and study their time evolution under the
dynamics given by (\ref{016}). If we choose appropriately peaked
states it should be possible to study to what extent the evolution
of wave packets follows the null geodesics of some cylindrical
spacetime metric and the length scales in which a metric gives an
accurate description of spacetime geometry.

Finally we want to point out that once the massless case is
understood it could be possible to use it as a guide to find a
consistent way to introduce other types of test fields (such as
massive scalars or electromagnetic fields) that further improve
our ability to explore quantum geometry and quantum gravity. This
will be the focus of our attention in the near future.

\begin{acknowledgments}
We want to thank G. Mena and J. M. Mart\'{\i}n Garc\'{\i}a for
drawing our attention to references \cite{Chandra,Wang} and also
them, A. Ashtekar, L. Garay, and M. Varadarajan for discussions.
IG is supported by a Spanish Ministry of Science and Education FPU
fellowship. This work is also supported by the Spanish MEC under
the research project BFM2002-04031-C02-02.
\end{acknowledgments}

\end{document}